\documentclass{amsart}
\date{February 16, 2007}

\def\const{\mathrm{const\;}}
\def\tr{\mathop{\mathrm{tr}}\nolimits} 

\def\cE{\mathcal{E}}

\def\gH{\mathfrak{H}}
\def\gQ{\mathfrak{Q}}

\def\rd{\mathrm{d}}

\newcommand{\cz}{\mathbb{C}} 
\newcommand{\nz}{\mathbb{N}} 
\newcommand{\rz}{\mathbb{R}} 

\newtheorem{theorem}{Theorem}[section]
\newtheorem{proposition}[theorem]{Proposition}
\newtheorem{lemma}[theorem]{Lemma}

\theoremstyle{definition}

\title[Relativistic Scott Correction]{The Ground State Energy of Heavy Atoms: Relativistic Lowering of the Leading Energy Correction}
\author[R. Frank]{Rupert L. Frank}
\address{Department of Mathematics\\ Royal Institute of Technology\\100
  44 Stockholm\\ Sweden}
\email{rupert@math.kth.se}

\author[H. Siedentop]{Heinz Siedentop}
\address{Ludwig-Maximilians-Universit\"at
  M\"unchen\\ The\-re\-sien\-stras\-se 39\\ 80333 M\"unchen\\ Germany}
\email{h.s@lmu.de}

\author[S. Warzel]{Simone Warzel}
\address{Department of Mathematics \\ Princeton
  University\\ Princeton, NJ 08544-1000\\USA}
\email{swarzel@princeton.edu}


 \thanks{We thank Elliott Lieb and Robert
  Seiringer for various supportive discussions. R.F. and H.S. thank
  the Departments of Mathematics and Physics of Princeton University
  for hospitality while this work was done. The work has been
  partially supported by the Swedish Foundation for International
  Cooperation in Research and Higher Education (STINT) (R.F.), the
  U.S. National Science Foundation, grant PHY 01 39984 (H.S.), and the
  Deutsche Forschungsgemeinschaft, grant SI 348/13-1 (H.S.).}

\begin{document}

\begin{abstract}
  We describe atoms by a pseudo-relativistic model that has its origin
  in the work of Chandrasekhar. We prove that the leading energy
  correction for heavy atoms, the Scott correction, exists. It turns
  out to be lower than in the non-relativistic description of
  atoms. Our proof is valid up to and including the critical coupling
  constant. It is based on a renormalization of the energy whose zero
  level we adjust to be the ground-state energy of the corresponding
  non-relativistic problem. This allows us to roll the proof back --
  by relatively simple technical means -- to results for the
  Schr\"odinger operator.
\end{abstract}
\maketitle

\section{Introduction}
\label{sec:1}

The energy of heavy atoms has attracted considerable interest that
dates back to the advent of quantum mechanics. As in classical
mechanics it soon became clear, that the exact solution of problems
involving more than two particles interacting through Coulomb forces
is not possible. Thomas \cite{Thomas1927} and Fermi
\cite{Fermi1927,Fermi1928} introduced their description of such atom
by the particle density and Lenz \cite{Lenz1932}, who wrote down the
corresponding energy functional which we will use here (see
\eqref{eq:15}), addressed this question and derived that the ground
state energy of atoms should decrease with the atomic number $Z$ as
$Z^{7/3}$. Scott predicted that this could be refined by an additive
$Z^2$-correction. Considerably later Schwinger \cite{Schwinger1980}
argued also for Scott's prediction. Schwinger \cite{Schwinger1981} and
Englert and Schwinger
\cite{EnglertSchwinger1984StatisticalAtom:H,EnglertSchwinger1984StatisticalAtom:S,EnglertSchwinger1985A}
even refined these considerations by adding more lower order terms
(see also Englert \cite{Englert1988}).

The challenge to address the underlying question whether the predicted
formulae would yield asymptotically correct results when compared with
the $N$-particle Schr\"odinger theory was for a long time
unsuccessful. It were Lieb and Simon who proved in their seminal paper
\cite{LiebSimon1977} that the prediction of Thomas, Fermi, and Lenz is
indeed asymptotically correct.  Alternative proofs were given by
Thirring \cite{Thirring1981} (lower bound), Lieb \cite{Lieb1981}, and
Balodis and Solovej \cite{MatesanzSolovej2000}. The Scott correction
was established by Hughes \cite{Hughes1986,Hughes1990} (lower bound),
and Siedentop and Weikard
\cite{SiedentopWeikard1987O,SiedentopWeikard1987U,SiedentopWeikard1988,SiedentopWeikard1989,SiedentopWeikard1991}
(lower and upper bound). In fact, even the existence of the
$Z^{5/3}$-correction conjectured by Schwinger was proven (Fefferman
and Seco
\cite{FeffermanSeco1989,FeffermanSeco1990,FeffermanSeco1990O,FeffermanSeco1992,FeffermanSeco1993,FeffermanSeco1994,FeffermanSeco1994T,FeffermanSeco1994Th,FeffermanSeco1995}).
Later these results were extended in various ways, e.g., the Scott
correction to ions (Bach \cite{Bach1989E,Bach1989}), to molecules
(Ivrii and Sigal \cite{IvriiSigal1993}, Solovej and Spitzer
\cite{SolovejSpitzer2003,SolovejSpitzer2003N}, Balodis
\cite{Balodis2004}), and to molecules in the presence of magnetic
fields (Sobolev \cite{Sobolev1994} and Ivrii \cite{Ivrii1999}). Ivrii
\cite{Ivrii1994} extended the validity of Schwinger's correction to
the molecular case.

Nevertheless, from a physical point of view, these considerations are
questionable, since large atoms force the bulk of the electrons on
orbits that are close to the nucleus (of order $Z^{-1/3}$) where the
electrons move with high speed which requires a relativistic
treatment. Schwinger \cite{Schwinger1981} has estimated this effect
concluding that they should contribute to the Scott correction wheras
the leading term should be unaffected by the change of model.  S\o
rensen \cite{Sorensen2005} was the first who proved that the
Thomas-Fermi term is indeed left unaffected when the non-relativistic
Hamiltonian is replace by the Chandrasekhar operator in the limit of
large $Z$ and large velocity of light $c$.  Cassanas and Siedentop
\cite{CassanasSiedentop2006} showed, that similarly to the
Chandrasekhar case, the leading energy is not affected for the
Brown-Ravenhall operator.

Recently, Solovej, S\o rensen, and Spitzer \cite{Solovej2006}
announced a proof that a correction is at most of the order $Z^2$
although no claim on the actual value of the coefficient was
made. (See also S\o rensen \cite{Sorensen1998} for the non-interacting
case).  In the present paper, we give an alternate proof of the Scott
correction of the Chandrasekar operator, which we present -- for
simplicity -- in the atomic case.  Our proof relies heavily on
semi-classical approximation for electrons that are far enough from
the nucleus. However, we use them only indirectly relying on known
results about the non-relativistic Scott correction. In addition we
use only relatively standard technical means as Lieb-Thirring and
Hardy inequalities. Our basic strategy is a renormalization of the
energy setting the energy of the Schr\"odinger atom as zero. Moreover,
we are able to extend the result of \cite{Solovej2006} to the case of
the critical coupling constant. In view of the corresponding situation
for the Dirac operator (see Remark \eqref{rem:dirac} after Theorem
\ref{t1}), this is a subtle and not at all obvious observation.

However, the question of whether the Schwinger correction which lives on the
scale $Z^{-2/3}$ also exists in this relativistic model and -- if so
-- cannot be answered with our techniques and is, therefore, left open.

The energy of an heavy atom is described
by a quadratic form
\begin{align}
  \label{eq:1}
    \cE^\# :\;  & \gQ_N \to \rz \notag \\ & \psi \mapsto
     \left\langle\psi,\left[\sum_{\nu=1}^N \left(T^\sharp - Z
     |x|^{-1}\right)_\nu + \sum_{1\leq\mu<\nu\leq N} |x_\mu-x_\nu|^{-1}\right]
     \psi\right\rangle
\end{align}
with
\begin{equation}
  \label{ee:1} 
  \gQ_N:=\bigwedge_{\nu=1}^N C_0^\infty(\rz^{3})\otimes \cz^{q} .
\end{equation}
The superscript $\#$ refers to the following two operators which are 
self-adjointly realized in $L^2(\rz^3)\otimes\cz^q$:
\begin{description}
\item[Chandrasekhar operator]
    \label{C}
    $T^C:=\sqrt{c^2p^2+c^4}-c^2$
\item[Schr\"odinger operator]
    \label{S}
    $T^S:= \tfrac12 p^2.$  
\end{description}  
The parameter $q\in \nz$ represents the possible number of spin states per
electron which -- physically -- has the value 2; $Z$ is the atomic
number, $c$ is the velocity of light, $N$ is the electron number.  
We use units in which $m=e^2=\hbar=1$.

A word about names: we address operators of the form $T^C+V$ with a
potential $V$ as Chandrasekhar operators, since the use of this kinetic
energy can be traced back at least to Chandrasekhar's semiclassical
treatment of the stability of stars \cite{Chandrasekhar1931} where it
can be viewed as the underlying operator. Later the use of $T^C$ has
been investigated by Weder \cite{Weder1975} and by Herbst
\cite{Herbst1977}. In the literature the operator is sometimes
addressed as pseudo-relativistic operator or Herbst operator.

In the following we assume that the system is neutral, i.e., $Z=N$, an
assumption that we make mainly because of notational convenience.  It
follows from Kato's inequality (with sharp constant),
$(2/\pi)|x|^{-1}\leq|p|$,
that the Chandrasekhar form $\cE^C$ is bounded from below, if and only if 
\begin{equation}
  \kappa:=Z/c \leq 2/\pi.
\end{equation}
Henceforth we assume this condition.

The ground state energy of a heavy atom with atomic number $Z$ is given by
\begin{equation}
  E^\#_{(\kappa)}(Z) := \inf\{\cE^\#(\psi)\,|\,\psi\in\gQ_N,\ \|\psi\|=1\}
\end{equation}
where $\#$ refers -- as above -- either to the Chandrasekhar or the
Schr\"odinger operator, the former being dependent additionally on
$\kappa$. We are interested in $E^C_\kappa(Z)$. However, $E^S(Z)$ will
also play an essential role, namely in regularizing the energy. Note
that $ \cE^C \leq \cE^S $, which implies that $E^C_\kappa(Z) \leq
E^S(Z)$. Our main result strengthens a result by Solovej, S\o rensen,
and Spitzer \cite{Solovej2006} in the atomic case to the critical
value of the coupling constant.

\begin{theorem}\label{t1}
  Let $\kappa\in(0,2/\pi]$ and $q\in\nz$. In the limit $Z\to\infty$
  with $\kappa = Z/c $ fixed and $N=Z$,
  \begin{equation}\label{eq:t1}
  E^C_\kappa(Z) = E^S(Z) - q s(\kappa) Z^2 + o(Z^2)
  \end{equation}
  where
  \begin{equation}
    \label{eq:scott}  
    s(\kappa) := \kappa^{-2}
    \tr\left[\left(\sqrt{p^2+1}-1-\kappa|x|^{-1}\right)_-
    -\left(\tfrac 12 p^2 - \kappa|x|^{-1}\right)_- \right] .
  \end{equation}
\end{theorem}

In \eqref{eq:scott} we used the notation $A_- :=
-A\chi_{(-\infty,0)}(A)$ for the negative part of a self-adjoint
operator $A$.

Several remarks apply:
\begin{enumerate}
\item As already mentioned in the introduction, the asymptotics of the
  ground-state energy $ E^S(Z) $ of the Schr\"odinger atom up to
  $o(Z^2)$ is given by the Thomas-Fermi energy and the Scott
  correction. To state this result precisely we introduce the
  Thomas-Fermi functional (Lenz \cite{Lenz1932})
\begin{equation}
  \label{eq:15}
  \cE_\mathrm{TF}(\rho):= \int_{\rz^3}\left[\frac35\gamma_\mathrm{TF}\rho(x)^{5/3} - \frac Z{|x|} \rho(x)\right]\,\rd x + D(\rho,\rho)
\end{equation}
where, in our units, $\gamma_\mathrm{TF}=(6\pi^2/q)^{2/3}/2$ and where
$$
D(\rho,\sigma):=\frac12\int_{\rz^3 \times\rz^3}{\rho(x)\sigma(y)\over
  |x-y|}\,\rd x \, \rd y 
$$ is the Coulomb scalar product. 
We define
\begin{equation}
  \label{eq:minimum}
  E_\mathrm{TF}(Z):= \inf\{\cE_\mathrm{TF}(\rho)\, |\, \rho\in
  L^1(\rz^3) \cap L^{5/3}(\rz^3),\ \rho\geq0 \}
\end{equation}
to be the minimal Thomas-Fermi energy. By scaling one finds that
$$E_\mathrm{TF}(Z)=E_\mathrm{TF}(1) Z^{7/3}.
$$ The asymptotic formula
\begin{equation}
E^S(Z) =  E_\mathrm{TF}(Z) + \tfrac 14 q Z^2 + O(Z^{47/24})
\end{equation}
was proven in \cite{SiedentopWeikard1987O,SiedentopWeikard1987U}; for
a lower bound only, see Hughes \cite{Hughes1986,Hughes1990}.
Inserting this into \eqref{eq:t1} one finds that
\begin{equation}
  E^C_\kappa(Z) = E_\mathrm{TF}(Z) + \left(\tfrac 14 - s(\kappa)\right) q Z^2 + o(Z^2).
\end{equation}
\item The spectral shift $s(\kappa)$ is monotone increasing with
  respect to $\kappa$ and strictly positive for $ \kappa > 0
  $. Indeed, by scaling $x\mapsto x/\kappa$,
  \begin{equation*}  
    s(\kappa) = 
    \tr\left[ \left(\sqrt{\kappa^{-2} p^2 +\kappa^{-4}} -\kappa^{-2} - |x|^{-1} \right)_- 
      -\left(\tfrac12 p^2 -|x|^{-1} \right)_-  \right],
  \end{equation*}
  and $\sqrt{\kappa^{-2} p^2 +\kappa^{-4}} -\kappa^{-2}$ is monotone
  decreasing with respect to $\kappa$.
\item\label{rem:dirac} It is part of our assertion that the operator
  in brackets in \eqref{eq:scott} belongs to the trace class.  In the
  subcritical case $\kappa<2/\pi$ this was already proved by
  S{\o}rensen \cite{Sorensen1998}. The finiteness of $ s(2/\pi) $
  should not be taken for granted:  in fact, when substituting the
  Dirac operator for the Chandrasekar operator in \eqref{eq:scott} it
  was shown numerically that the corresponding spectral shift diverges
  at the critical coupling \cite{Sorensen1998}.
\end{enumerate}

Since neither the Schr\"odinger nor the Chandrasekhar operator depend
explicitly on spin, we shall assume henceforth $q=1$; the general case
follows along the same line. We prove Theorem \ref{t1} in Section
\ref{sec:2} after having established a precise bound on the spectral
shift for one-particle operators in the next section.


\section{Bound on the Spectral Shift}\label{sec:3}

For any real-valued potential $v$ for which the following operators can be defined
according to Friedrichs, we set
\begin{align}
   S(v) & := \tfrac12p^2 - v, \label{eq:s}\\ 
   C(v) & :=\sqrt{p^2 + 1} -1 - v, \label{eq:c}
\end{align}
the Schr\"odinger respectively Chandrasekar operator in $L^2(\rz^3)$.
We assume $c=1$ throughout this section.

If the potential $v$ is radially symmetric, both the Schr\"odinger
and the Chandrasekhar operator commute with the angular momentum
operators allowing for a decomposition into the corresponding invariant
subspaces. For each $l\in\nz_0$ the subspace $\gH_l$
spanned by the spherical harmonics $Y_{l,m} $ with $m=-l,\ldots,l$, is
an invariant subspace of $S(v)$ and $C(v)$, and
$\oplus_{l=0}^\infty \, \gH_l = L^2(\rz^3)$.  We write $\Lambda_l$ for
the orthogonal projection onto $\gH_l$ and
\begin{equation}
  \tr_l(A) := \tr(\Lambda_l A)
\end{equation}
for the corresponding reduced trace. 

Our main result in this section concerns the decay of the spectral
shift 
$$\tr_l\left(\left[ C(v) \right]_- -\left[S(v)\right]_- \right)$$
as the angular momentum $l$ increases. We shall prove
\begin{theorem}
  \label{t:3}
  There exists a constant $M$ such that for all $ \mu \geq 0 $ and for
  all $ l \in \mathbb{N}_0 $ and for all $v:[0,\infty)\to[0,\infty)$ satisfying
      \begin{equation}\label{eq:vbound}
	v(r) \leq \tfrac2\pi \, r^{-1}
      \end{equation}
       the sum of eigenvalue differences for angular momentum $l$ is
      bounded according to
      \begin{equation}\label{eq:t3}
	0 \leq \tr_l \left(\left[ C(v) + \mu \right]_- - \left[S(v) +
	\mu \right]_- \right) \leq M (l+1)^{-2} .
      \end{equation}
\end{theorem}

This theorem shows that there is an effective cancelation in the
difference in \eqref{eq:t3}. Indeed, if $v(r)=\kappa r^{-1}$, then
\begin{equation*}
  \tr_l \left[S(\kappa r^{-1})\right]_- = (2l+1) \frac{\kappa^2} 2
  \sum_{n=1}^\infty \frac1{(n+l)^2}
\end{equation*}
and this does not decay at all as $l\to\infty$. We note also that
\eqref{eq:t3} implies that the operator $\left(\sqrt{p^2 +1} -1
-\kappa|x|^{-1}\right)_- -\left(\tfrac12 p^2-\kappa|x|^{-1}\right)_-$
appearing in Theorem \ref{t1} is trace class for any
$\kappa\in(0,\frac2\pi]$.


\subsection{Reminder on Lieb-Thirring Estimates}

In the proof of Theorem \ref{t:3} we use the following relativistic
Lieb-Thirring inequalities due to Daubechies \cite{Daubechies1983}.

\begin{proposition}\label{prop:Daubechies}
  For any $ \gamma > \tfrac{1}{2} $ there exists a constant $L_\gamma$ such that
  for all $ l \geq 0 $
\begin{equation}\label{eq:daubechies}
  \tr_l \left[ C(v) \right]_-^\gamma \leq L_\gamma (2l+1) \int_0^\infty \left[
    [v(r)]_+^{1+\gamma} + [v(r)]_+^{\tfrac{1}{2} +\gamma} \right] \rd r.
\end{equation}
\end{proposition}

Proposition \ref{prop:Daubechies} is also valid for $\gamma=\tfrac12$,
but we will not need this fact.

\begin{proof}
Since $ \tr_l \left[ C(v) \right]_-^\gamma \leq (2l+1) \tr_0 \left[
C(v) \right]_-^\gamma $, it suffices to verify the claim for $ l = 0
$. If we extend $v$ to an even function $\tilde v$ on $\mathbb{R}$,
then $C(v)$ is unitarily equivalent to the part of the
\emph{whole-line} operator $\sqrt{p^2+1}-1-\tilde v$ on antisymmetric
functions. In the whole-line case, the result follows by evaluating
the integral in \cite[Eq.~(2.14)]{Daubechies1983}.
\end{proof}

Our treatment of the critical case $\kappa=\frac2\pi$ is based on the
following inequality \cite[Theorem 11]{LiebYau1988} of Lieb and Yau.
\begin{proposition}\label{liebyau}
Let $I$ be a function with support in $\{x\in\rz^3 : |x|\leq 1\}$. Then for all $\mu>0$
\begin{equation*}
\tr \left[ I\left(|p|-\tfrac2\pi|x|^{-1}- \mu\right)\overline I \right]_- 
\leq \const \mu^4 \int |I(x)|^2 \rd x.
\end{equation*}
\end{proposition}


\subsection{Finiteness of Partial Traces}

In \eqref{eq:t3} appears the trace of the difference of the operators
$\left[ C(v) + \mu \right]_-$ and $\left[S(v) + \mu \right]_-$. We
begin by proving that both operators separately have finite
traces. Since $S(v)\leq C(v)$ (see also \eqref{eq:banff} below) it
suffices to prove this in the relativistic case.

\begin{lemma}\label{t:3.2}
  For all $l\in\nz_0$ one has $\tr_l \left[C\left(\tfrac2\pi
    |x|^{-1}\right)\right]_-<\infty$.
\end{lemma}
\begin{proof}
  Pick a Lipschitz function $\varphi:\rz_+\rightarrow [0,\pi/2]$ with
  Lipschitz constant $ \phi_0 $ which vanishes for $r\leq 1/2$ and
  which is $\pi/2$ for $r\geq 1$. Then $I:=\cos(\varphi)$ has compact
  support around the origin and, furthermore, it constitutes together
  with $A:=\sin(\varphi)$ a quadratic partition of unity, i.e.,
  $I^2+A^2=1$. According to Lieb and Yau \cite[Theorem 9]{LiebYau1988}
  we have the localization formula
  \begin{equation}
    \label{eq:3.3}
    \langle \psi,(p^2+1)^{1/2}\psi \rangle = \langle I \psi
    ,(p^2+1)^{1/2} I \psi \rangle + \langle A \psi ,(p^2+1)^{1/2}A\psi
    \rangle - \langle \psi,L\psi \rangle
  \end{equation}
  for $\psi \in L^2(\rz^3)$. Here $L$ is the bounded integral operator
  on $ L^2(\rz^3) $ with non-negative kernel given in terms of a
  Bessel function
  \begin{equation}
    \label{eq:3.4}
    L(x,y):= K_2(|x-y|) 
    {\sin^2\left[(\varphi(|x|)-\varphi(|y|))/2\right]\over\pi^2|x-y|^2}.
  \end{equation}
  We shall estimate this localization error by a multiplication
  operator. More precisely, we shall show that there exists a constant
  $M>0$ such that
  \begin{equation}\label{eq:locpot}
  \langle\psi,L\psi\rangle \leq	M \langle\psi,e^{-|x|}\psi\rangle.
  \end{equation}
  To prove this, we note that by the Schwarz inequality we have 
  \begin{equation}
    \label{eq:3.5}
    \begin{split}
      \langle\psi,L\psi\rangle \leq & \int_{\rz^3 }\rd x\;  |\psi(x)|^2
      \int_{\rz^3}\rd y\; K_2(|x-y|)
      {\sin^2((\varphi(|x|)-\varphi(|y|))/2)\over\pi^2|x-y|^2} \\ 
      \leq & \left(\frac{\phi_0}{2\pi}\right)^2 \int_{|x|<1}\rd x\;
      |\psi(x)|^2 \int_{\rz^3} \rd y \;K_2(|x-y|) \\ & +
      \left(\frac{\phi_0}{2\pi}\right)^2 \int_{|x|\geq 1}\rd x\;
      |\psi(x)|^2 \int_{|y|< 1} \rd y\; K_2(|x-y|)\\
      =: & \langle\psi,v_I\psi\rangle + \langle\psi,v_A\psi\rangle.
      \end{split}
  \end{equation}  
  Since $\int_0^\infty \rd r  r^2 K_2(r)= 3\pi/2$ \cite[Formula
  11.4.22]{Luke1968} we have
  \begin{equation*}
  	v_I(x):= \left(\frac{\phi_0}{2\pi}\right)^2
  	\chi_{\{|x|<1\}}(x) \int_{\rz^3}\rd y \, K_2(|x-y|) =
  	\frac{3\phi_0^2}{2} \chi_{\{|x|<1\}}(x).
  \end{equation*}	
  Moreover, since $K_2(r)= 2/r^2+O(1)$ as $ r \downarrow
  0 $ and $ K_2(r)\sim\sqrt{\pi/(2r)}\exp(-r)$ as $r\to\infty$
  \cite{Olver1968}, the function
  \begin{equation*}
  v_A(x) := \left(\frac{\phi_0}{2\pi}\right)^2 \chi_{\{|x|\geq 1\}}(x) \int_{|x+y|<1}\rd
  y\, K_2(|y|)
  \end{equation*}
  is well-defined and satisfies $v_A(x)\leq \const e^{-|x|}$. This
  proves \eqref{eq:locpot}.
  
  Combining \eqref{eq:3.3} and \eqref{eq:locpot} we find that
  \begin{multline}
    \tr_l \left[C\left(\tfrac2\pi |x|^{-1}\right)\right]_-\\ \leq
    \tr_l \left[ I \left( C\left(\tfrac{2}{\pi}|x |^{-1} + M
    e^{-|x|}\right)\right) I\right]_- + \tr_l \left[ A \left(
    C\left(\tfrac{2}{\pi}|x |^{-1} + M e^{-|x|}\right)\right) A
    \right]_-.
  \end{multline}
  To estimate the inner part we use that
  $$ I \left( C\left(\tfrac{2}{\pi}|x |^{-1} + M
  e^{-|x|}\right)\right) I \geq I \left( |p| - \tfrac{2}{\pi}|x |^{-1}
  - 1 - M e^{-1} \right) I. $$ It follows therefore from Proposition
  \ref{liebyau} that the corresponding trace is finite (even when
  summed over all $l$). For the outer part we use
  $$ A \left( C\left(\tfrac{2}{\pi}|x |^{-1} + M e^{-|x|}\right)\right) A
  \geq C\left(\chi_{\{|x|\geq\tfrac12\}} \tfrac{2}{\pi}|x |^{-1} + M e^{-|x|} \right).
  $$
  The corresponding trace is finite by Proposition \ref{prop:Daubechies}.
\end{proof}


\subsection{Angular Momentum Barrier Inequalities\label{app:b}}
A straightforward consequence of Hardy's inequality, which we will frequently exploit, is 
\begin{lemma}\label{hardy}
Let $l\in\nz_0$. Then, as operators in $\gH_l$
\begin{equation}
  \label{eq:hardy}
  p^2 \geq (l + \tfrac{1}{2})^2 r^{-2}.
\end{equation}
\end{lemma}
\begin{proof}
Writing the Laplacian in spherical coordinates we find that $p^2$ in
$\mathfrak H_l$ is unitarily equivalent to $p_r^2+l(l+1)r^{-2}$ in
$L^2(\mathbb{R}_+)$. The claim follows hence from Hardy's inequality,
$p_r^2\geq (2r)^{-2}$.
\end{proof}

By operator monotonicity of the square root, \eqref{eq:hardy}
implies the (non-sharp) inequality $|p| \geq (l + \tfrac{1}{2})
r^{-1}$ in $\mathfrak H_l$. We shall need an analogue of this
inequality for the operator $C(0)$ instead of $|p|$. Note that
$\sqrt{p^2+1}-1$ behaves as $\tfrac12 p^2$ for small $p$. Since `small
$p$' corresponds intuitively to `large $r$', we cannot expect that
$C(0)$ controls an $r^{-1}$ decay. But it \emph{does} control an
$r^{-1}$ singularity. This is the content of
\begin{lemma}\label{ambi} 
  Let $l\in\nz_0$, $R>0$ and $M_l(R) :=
  (l+\tfrac12)^2/\left(R+\sqrt{R^2+(l+\tfrac12)^2}\right)$. Then, as
  operators in $\gH_l$
  \begin{equation}
    C(0) \geq M_l(R)\, \chi_{\{r\leq R\}}(r) \, r^{-1}.
  \end{equation}
\end{lemma}
\begin{proof}
The inequality \eqref{eq:hardy} and operator monotonicity of the
square root imply in $\gH_l$
$$ 
\sqrt{p^2 + 1 } - 1 \geq \sqrt{(l + \tfrac{1}{2})^2 \, r^{-2} + 1} - 1;
$$
the claim follows by determining the solution of the equation  
$$ \sqrt{(l + \tfrac{1}{2})^2 \, r^{-2} + 1} = 1+ M r^{-1}.$$
\end{proof}

The core of Theorem \ref{t:3} is contained in the following
\begin{lemma}
  \label{sshift}
  There exists a constant such that for all $ v: [0,\infty) \to
  [0,\infty) $ satisfying \eqref{eq:vbound} for all $ \mu \geq 0 $ and
  for all $ l \in \mathbb{N} $ one has
  \begin{equation}\label{eq:sshift}
    \begin{split}
      0 & \leq \tr_l \left(\left[ C(v) + \mu \right]_- - \left[S(v) +
      \mu \right]_- \right) \\ & \leq \const \left( \tr_l \left[
      C(w_l)\right]_-^2 + (l+\tfrac12)^{-2} \tr_l \left[
      C(w_l)\right]_- \right)
    \end{split}
  \end{equation}
  where $w_l(r) := 10 \, r^{-1} \chi_{\{r\geq l^2/4\}}(r)$.
\end{lemma}
\begin{proof}
The identity 
\begin{equation}
  \label{eq:banff}
  \tfrac12 p^2 = C(0) + \tfrac12 C(0)^2 
\end{equation}
implies the non-negativity asserted in \eqref{eq:sshift}.

To prove the second inequality in \eqref{eq:sshift} we shall first
assume (in addition to \eqref{eq:vbound}) that $v$ is a bounded
function and that $\mu>0$. Once the inequality is proved in this case
(with a constant independent of $\mu$ and the supremum of $v$), we can
apply it to the cut-off potential $v_M:=\min\{v,M\}$. 

By monotone convergence $C(v_M)$ and $S(v_M)$ converge to $C(v)$ and
$S(v)$ in strong resolvent sense \cite[Thm.~1.2.3]{Davies1990}, and
therefore \cite[Thm.~VIII.20]{ReedSimon1972},
\cite[Thm.~2.7]{Simon2005} for any $\mu>0$, $\liminf_{M\to\infty}
\tr_l \left[ C(v_M) + \mu \right]_- \geq \tr_l \left[ C(v) + \mu
\right]_-$ and similarly for $S(v_M)$. But the reverse inequalities
are also true, since $C(v_M)\geq C(v)$ and $S(v_M)\geq S(v)$. Hence we
conclude that $\tr_l \left(\left[ C(v_M) + \mu \right]_- -
\left[S(v_M) + \mu \right]_- \right)$ converges to the corresponding
quantity with $v_M$ replaced by $v$. Finally, we can use
Lemma~\ref{t:3.2} to extend the result to $\mu\to 0$.

Thus we may assume $v$ to be bounded, $\mu>0$ and denote by $ \gamma_l
$ the orthogonal projection onto the eigenspace of $C(v)$
corresponding to angular momentum $l$ and eigenvalues less or equal
than $-\mu$. Since $v$ is bounded, any eigenfunction of $C(v)$ lies in
the form domain of $S(v)$. Hence the variational principle together
with \eqref{eq:banff} yields
\begin{equation}\label{eq:varprinc}
  2 \tr_l \left( \left[C(v) + \mu \right]_- - \left[S(v) + \mu \right]_- \right)
   \leq \tr_l \left[ C(0)^2  \gamma_l \right] .
\end{equation}
Again the boundedness of $v$ and the finite rank of $\gamma_l$ imply
that $\tr_l \left[ C(0)^2 \gamma_l \right]$ is finite. Using the
eigenvalue equation and the bound \eqref{eq:vbound} on the potential
we estimate this term further as follows.
\begin{equation}
  \label{eq:4.1}
 \tr_l \left[ C(0)^2 \gamma_l \right] 
 \leq \tr_l \left[ C(v) \right]_-^2 + \tr_l \left[ v^2 \gamma_l \right]
 \leq \tr_l \left[ C(\tfrac2\pi |x|^{-1})\right]_-^2 
 + (\tfrac2\pi)^2 \tr_l \left[ |x|^{-2} \gamma_l \right] .
\end{equation}
The last term in the above inequality is bounded using \eqref{eq:hardy} and \eqref{eq:banff},
\begin{equation}
  \label{eq:4.2}
  \left\|  |x|^{-1} \psi_l \right\|^2  
   \leq \left(l+\tfrac12\right)^{-2}  \left\| |p| \psi_l \right\|^2    
    =  \left(l+ \tfrac12\right)^{-2} \left(\left\| C(0)   \psi_l \right\|^2 
   + 2 \langle \psi_l , C(0)  \psi_l \rangle \right) 
\end{equation}
valid for $ \psi_l \in \gH_l $. Since $ l \geq 1 $ we have
$(\tfrac2\pi)^2(l+\tfrac12)^{-2}\leq\tfrac12$ and thus the last two
estimates may be summarized as
\begin{equation}
   \tr_l \left[ C(0)^2 \gamma_l \right] \leq 2 \tr_l \left[
	C(\tfrac2\pi |x|^{-1})\right]_-^2 + 4 (\tfrac2\pi)^2 (l +
	\tfrac12)^{-2} \tr_l \left[ C(0) \gamma_l \right] .
\end{equation}
In view of \eqref{eq:varprinc} the assertion will follow, if we can prove
\begin{equation}
  \label{eq:pulltooth}
  \tr_l \left[ C(\tfrac2\pi |x|^{-1})\right]_-^2  \leq \tr_l \left[ C(w_l)\right]_-^2, 
  \qquad
  \tr_l \left[ C(0)  \gamma_l \right] \leq \tr_l \left[ C(w_l)\right]_-.
\end{equation}
We begin with the (more difficult) second inequality. We have
\begin{equation}
  \tr_l \left[ C(0) \gamma_l \right] \leq \tr_l \left[ v \gamma_l
  \right] \leq \tfrac2\pi \tr_l \left[ |x|^{-1} \gamma_l \right] \leq
  \tr_l \left[ |x|^{-1} \gamma_l \right].
\end{equation}
We apply Lemma~\ref{ambi} with $ R=l^2/4 $ to bound the last
term. Since $M_l(l^2/4) \geq 5/4$ for $l\geq 1$ we obtain
\begin{equation}
  \label{eq:fifth}
   \left\langle \psi_l , |x|^{-1} \psi_l \right\rangle 
   \leq \frac{4}{5} \left\langle \psi_l , C(0) \psi_l \right\rangle
   + \left\langle \psi_l , \chi_{\{ r \geq l^2/4\}}(|x|)
   \, |x|^{-1} \psi_l \right\rangle.
\end{equation}
The last two estimates can be summarized as
\begin{equation}
 	\tr_l \left[ C(0) \gamma_l \right] \leq - \tr_l \left[ C(w_l)
		\gamma_l \right] \leq \tr_l \left[ C(w_l) \right]_-,
\end{equation}
which proves the second inequality in \eqref{eq:pulltooth}. We proceed
similarly to prove the first one. Indeed, by \eqref{eq:fifth}
\begin{equation}
  C(\tfrac2\pi |x|^{-1}) 
  \geq \tfrac15 C(0) + \chi_{\{ r \leq l^2/4\}}(|x|) \, |x|^{-1} - \tfrac2\pi |x|^{-1}
  \geq \tfrac15 C(w_l) 
\end{equation}
and hence $\tr_l\left[C(\tfrac2\pi |x|^{-1})\right]_-^2 \leq
\tfrac1{25}\tr_l\left[C(w_l)\right]_-^2$. This completes the proof of
the lemma.
\end{proof}

Now everything is in place for the 
\begin{proof}[Proof of Theorem \ref{t:3}]
The boundedness of the trace in \eqref{eq:t3} for $ l = 0 $ is implied
by Lemma~\ref{t:3.2} below, and its non-negativity follows from
\eqref{eq:banff}.  For $l\geq 1$ we use Lemma \ref{sshift} and note
that
\begin{equation}
\tr_l \left[ C(w_l)\right]_-^2 \leq \const l^{-2},
\qquad
\tr_l \left[ C(w_l)\right]_-^2 \leq \const
\end{equation}
by Proposition \ref{prop:Daubechies}.
\end{proof}


\section{Proof of the Main Results: Renormalization of the
    Relativistic Operator\label{sec:2}}

The strategy for our proof of the main results is to use the
Schr\"o\-din\-ger operator as a regularization for the relativistic
problem, i.e., we will use it to eliminate the main contribution to
the energy -- the Thomas-Fermi energy -- and to focus only on the
energy shift of the low lying states where the electron-electron
interaction plays no role and the unscreened problem remains.

Recall that Theorem \ref{t1} for $q=1 $ reads
\begin{equation}
  \lim_{Z\to\infty} {E^S(Z) - E^C_\kappa(Z) \over Z^2}= s(\kappa).
\end{equation}
We will show this claim in two steps, namely that the upper limit
and the lower limits exist and are given be the same expression,
namely the coefficient of the $Z^2$-correction claimed in the theorem.
That this coefficient is finite was already remarked after Theorem \ref{t:3}.

\subsection{Upper Bound on the Energy Difference -- Lower Bound on the
  Relativistic Energy}

Lieb and Simon \cite{LiebSimon1977} showed that the Thomas-Fermi minimization
problem \eqref{eq:minimum} has a unique minimizer $\rho_Z$, the
Thomas-Fermi density. It fulfills the scaling relation
\begin{equation}
  \label{skalierung}
  \rho_Z(x):= Z^2\rho_1(Z^{1/3}x).
\end{equation}
We define the radius of the Thomas-Fermi exchange hole at point $x\in\rz^3$ as
the unique minimal radius $R_Z(x)$ for which
\begin{equation}
  \int_{|x-y|\leq R_Z(x)} \rho_Z(y) \rd y = \frac 12.
\end{equation} 
We denote the exchange-hole-reduced Thomas-Fermi screening potential by
\begin{equation}
  \label{eq:potex}
  \chi_\mathrm{TF}(x):= \int_{|x-y|>R_Z(x)}{\rho_Z(y)\over|x-y|}\rd y
\end{equation}
and the corresponding one-particle operators by
\begin{align}
  \label{eq:stf}
  S_\mathrm{TF} &= S(Z|x|^{-1} - \chi_\mathrm{TF})\\
  \label{eq:ctf}
  C_\mathrm{TF} &= C_c(Z|x|^{-1}-\chi_\mathrm{TF})
\end{align}
both self-adjointly realized in $L^2(\rz^3)$. Here we use a notation
similar to that in Section~\ref{sec:3},
\begin{equation}
   C_c(v)  := \sqrt{p^2 c^2 + c^4} -c^2 - v.
\end{equation}
We remark that we slightly deviate from the more usual choice
$Z|x|^{-1}-\rho_Z*|\cdot|^{-1}(x)$ for the screened potential. This is
motivated by the correlation inequality \eqref{eq:correlation} below.
The concept of an exchange hole can be traced back to Slater
\cite{Slater1951}. It also has been used to estimate the
exchange-correlation energy (Lieb \cite{Lieb1979}, Lieb and Oxford
\cite{LiebOxford1981}).

We shall express the many-particle ground-state energy in terms of
quantities involving the above one-particle operators. In the
relativistic case we use the correlation inequality of
\cite{Mancasetal2004} to obtain a lower bound on the many-particle
ground-state energy.

\begin{lemma}\label{lemma:chlower}
For all $ L \in \mathbb{N} $, 
\begin{equation}\label{eq:chlower}
  E^C_\kappa(Z) \geq -\sum_{l=0}^{L-1} \tr_l \left[ C_c(Z|x|^{-1}) \right]_- 
  - \sum_{l=L}^{\infty} \tr_l \left[ C_{\mathrm{TF}}\right]_- - D(\rho_Z,\rho_Z) .
\end{equation}
\end{lemma}
\begin{proof}
We use the correlation inequality \cite[Equation (14)]{Mancasetal2004}
\begin{equation}\label{eq:correlation}
  \sum_{1\leq\mu<\nu\leq N} |x_\mu-x_\nu|^{-1} \geq \sum_{\nu=1}^N
  \chi_\mathrm{TF}(x_\nu) - D(\rho_Z,\rho_Z),
\end{equation}
to bound $ E^C_\kappa(Z) $ from below by the ground-state energy of 
$$\sum_{\nu=1}^N \big( C_{\mathrm{TF}} \big)_\nu - D(\rho_Z,\rho_Z).$$ 
This yields
\begin{equation}
  E^C_\kappa(Z) \geq -\tr [C_\mathrm{TF} ]_- -
  D(\rho_Z,\rho_Z).
\end{equation}
We split the trace according to angular momentum and use the operator
inequality $ C_\mathrm{TF} \geq C_c(Z|x|^{-1}) $ for all $ l \leq L -1
$ to obtain the assertion.
\end{proof}

In the non-relativistic case, we recall
\begin{proposition}\label{prop:schrupper}
Let $ L := \big[ Z^{1/9} \big] $. Then, as $ Z \to \infty $, 
\begin{equation}\label{eq:schrupper}
  E^S(Z) = -\sum_{l=0}^{L-1} \tr_l \left[ S(Z|x|^{-1}) \right]_- 
  - \sum_{l=L}^{\infty} \tr_l \left[ S_{\mathrm{TF}}\right]_- 
  - D(\rho_Z,\rho_Z) + O(Z^{47/24}).
\end{equation}
\end{proposition}

\begin{proof}
The same argument as in Lemma \ref{lemma:chlower} yields the lower bound
\begin{equation}\label{eq:schlower1}
E^S(Z) \geq -\sum_{l=0}^{L-1} \tr_l \left[ S(Z|x|^{-1}) \right]_- -
\sum_{l=L}^{\infty} \tr_l \left[ S_{\mathrm{TF}}\right]_- -
D(\rho_Z,\rho_Z) .
\end{equation}
Note that the $\chi_{TF}\leq\rho_Z*|\cdot|^{-1}$. Hence \cite[Theorem
1]{SiedentopWeikard1989} and the proof of this theorem (in particular,
\cite[Lemma 2]{SiedentopWeikard1989}, see also
\cite{SiedentopWeikard1991}) show that one can further estimate
\begin{equation}\label{eq:schlower2}
\begin{split}
  & -\sum_{l=0}^{L-1} \tr_l \left[ S(Z|x|^{-1}) \right]_- 
  - \sum_{l=L}^{\infty} \tr_l \left[ S_{\mathrm{TF}}\right]_- 
  - D(\rho_Z,\rho_Z) \\
  & \qquad\qquad \geq E_\mathrm{TF}(Z) + \tfrac 14 Z^2 - \const Z^{17/9}\log Z.
\end{split}
\end{equation}
On the other hand, one has the upper bound \cite[Lemmas~3.1 and
4.1]{SiedentopWeikard1987O}
\begin{equation}
E^S(Z)\leq E_\mathrm{TF}(Z) + \tfrac 14 Z^2 +\const Z^{47/24}
\end{equation}
Combining this with \eqref{eq:schlower1} and \eqref{eq:schlower2} we obtain the assertion.
\end{proof}

\begin{proof}[Proof of Theorem~\ref{t1} -- first part]
Choosing $ L= \big[ Z^{1/9} \big] $ and combining
Lemma~\ref{lemma:chlower} and Proposition~\ref{prop:schrupper} we
obtain
\begin{multline}
  \label{eq:U1}
  E^S(Z)-E^C_\kappa(Z) \leq -\sum_{l=0}^{L-1} \tr_l
  \left(\left[S(Z|x|^{-1})\right]_-
  -\left[C_c(Z|x|^{-1})\right]_-\right) \\ - \sum_{l=L}^\infty \tr_l
  \left( \left[S_\mathrm{TF}\right]_-
  -\left[C_\mathrm{TF}\right]_-\right) + \const Z^{47/24}.
\end{multline}
We note that by scaling $x\mapsto x/c$, the operators $S_\mathrm{TF}$
and $C_\mathrm{TF}$ are unitarily equivalent to the operators $
\kappa^{-2} Z^2 S(\kappa|x|^{-1}-\chi_Z) $ and $ \kappa^{-2} Z^2
C_1(\kappa|x|^{-1}-\chi_Z) $, both acting in $ L^2(\mathbb{R}^3) $,
where
\begin{equation}
  \label{eq:chitilde}
  \chi_Z(x):=\kappa^2 Z^{-2}\chi_\mathrm{TF}(\kappa x/Z).
\end{equation}
This implies
\begin{equation}
  \label{eq:U2}
  \limsup_{Z\to\infty}{E^S(Z)-E^C_\kappa(Z)\over Z^2}  \leq
  \kappa^{-2}\limsup_{Z\to\infty} \left( \Sigma_1(Z) + \Sigma_2(Z)
  \right)
\end{equation}
where
\begin{align*}
   \Sigma_1(Z) := & \sum_{l=0}^{L-1} \tr_l
  \left(\left[C_1(\kappa|x|^{-1})\right]_- - \left[S(\kappa|x|^{-1})\right]_-\right)  \\ 
    \Sigma_2(Z) := & \sum_{l=L}^\infty
  \tr_l \left(\left[C_1(\kappa|x|^{-1}-\chi_Z)\right]_- - \left[S(\kappa|x|^{-1}-\chi_Z\right]_-\right) .
\end{align*}
Theorem~\ref{t:3} implies that the summands in both sums on the
right-hand side are non-negative and bounded by $\const(l+1)^{-2}$
independently of $ Z $.  Therefore the first sum actually converges
\begin{equation}
  \label{U3}
  \limsup_{Z \to \infty} \Sigma_1(Z) = \sum_{l=0}^{\infty}
  \tr_l \left(\left[C_1(\kappa|x|^{-1})\right]_- - \left[S(\kappa|x|^{-1})\right]_-\right).
\end{equation}
Moreover, the second sum converges to zero,
\begin{equation}
\limsup_{Z \to \infty}  \Sigma_2(Z) \leq \const \, \limsup_{Z \to \infty} \sum_{l=L}^\infty (l+1)^{-2} = 0 .
\end{equation}
This concludes the proof of the upper bound on the energy difference.
\end{proof}

We remark that $\Sigma_2(Z)\leq\const Z^{-1/9}$, hence we have actually shown that
\begin{equation}
E^S(Z)-E^C_\kappa(Z) \leq s(\kappa)Z^2 + \const Z^{47/24}.
\end{equation}


\subsection{Lower Bound on the Energy Difference -- Upper Bound on the
  Relativistic Energy}

Following \cite{SiedentopWeikard1987O} we define two one-particle
density matrices $d^S$ and $d^C$ as sums
\begin{equation}
  \label{eq:d}
  d^\# = \sum_{l=0}^\infty d^\#_{l}. 
\end{equation}
As above, we use the convention that $\#$ refers either to the
Schr\"o\-dinger case or to the Chandrasekhar case. The operators
$d^\#_{l}$ are defined in $ L^2(\mathbb{R}^3) $ through their integral
kernels
\begin{equation}
  \label{eq:dl}
  d^\#_{l}(x,y) := \sum_{n=1}^\infty w_{n,l}\sum_{m=-l}^l \psi_{n,l,m}^\#(x) \overline{\psi_{n,l,m}^\#(y)} .
\end{equation}
The weights $w_{n,l}$ and the functions
$\psi^\#_{n,l,m}$ are defined separately for angular momentum $ l < L $ and $ l \geq L $ where
$ L $ will be chosen later in a $ Z $-dependent way.

\begin{description}
\item[Case $ l < L $] We define $ \psi^\#_{n,l,m} $ as the $n$-th
  eigenfunction of $ S(Z|x|^{-1}) $ restricted to angular momentum
  $(l,m)$, respectively of $ C_c(Z|x|^{-1})$ restricted to angular
  momentum $ (l,m) $, with the normalization $\|\psi^\#_{n,l,m}\|_2
  =1$. Note that this function is of the form $ \psi^\#_{n,l,m}(x) =
  \varphi_{n,l}(|x|) Y_{l,m}(x/|x|) $ with a radial function $
  \varphi_{n,l} $.  The weights $w_{n,l}$ are defined independently of
  $m$ by
  \begin{equation}
    w_{n,l} :=
    \begin{cases} 
      1 & n\leq K-l,\\
      0 & n> K-l.
    \end{cases}
  \end{equation}
  where $K :=[dZ^{1/3}]$ with $d$ some
  positive constant independent of $Z$.
\item[Case $ l \geq L $] We choose $ \psi^\#_{n,l,m}(x) =
\varphi_{n,l}(|x|) Y_{l,m}(x/|x|) $ where the functions
$\varphi_{n,l}$, as well as the weights $w_{n,l}$, are defined exactly
as in \cite[Section 2]{SiedentopWeikard1987O} independently of $ \# $.
(The exact form of the functions and the values of the weights for
$l\geq L$ are irrelevant in our context.)
\end{description}

Note that the above construction guarantes $d^\#$ to be density
matrices, i.e., $0\leq d^\# \leq 1$. Moreover, by the choice of $L$,
$K$, and $w_{n,l}$ one can assure that $\tr d^\#\leq Z$. (For $\#=S$
this is proved in \cite[Corollary~4.1]{SiedentopWeikard1987O}, and
follows hence also for $\#=C$.)

Since $d_l^\#$ is independent of $\#$ for $l\geq L$ we drop the
superscript in this case. Moreover, we shall use the notations
\begin{equation*}
d^\#_< := \sum_{l=0}^{L-1} d^\#_{l},
\quad d_> := \sum_{l=L}^\infty d_{l},
\end{equation*}
and
\begin{equation*}
\rho^\#_l(x) := d^\#_l(x,x),
\quad \rho^\#_<(x) := d^\#_<(x,x), 
\quad \rho_>(x) := d_>(x,x).
\end{equation*}

We recall now that the density matrix $d^S$ gives an energy which is
correct up to the order we are interested in. More precisely, one has
\begin{proposition}\label{prop:schlower}
Let $ L := \big[ Z^{1/12} \big] $. Then, for sufficiently large $Z$,
\begin{equation}
  E^S(Z) = \tr[S(Z|x|^{-1}) d^S] + D(\rho^S,\rho^S) + O(Z^{47/24}).
\end{equation}
\end{proposition}

\begin{proof}
It is shown in \cite{SiedentopWeikard1987O} that for sufficiently large $Z$,
\begin{equation*}
E^S(Z) \leq \tr[S(Z|x|^{-1}) d^S] + D(\rho^S,\rho^S)
\leq E_\mathrm{TF}(Z) + \tfrac 14 Z^2 +\const Z^{47/24} .
\end{equation*}
Combining this with the lower bound on $E^S(Z)$ which was recalled in
\eqref{eq:schlower1} and \eqref{eq:schlower2}, we obtain the
assertion.
\end{proof}

We decrease the ground state energy further by dropping a part of the Coulomb energy,
\begin{equation}\label{eq:schlower}
  E^S(Z) \geq \tr[S(Z|x|^{-1}) d^S_<]
  + \tr[S(Z|x|^{-1}) d_>] +D(\rho_>,\rho_>) - \const Z^{47/24}.
\end{equation}
For an upper bound in the relativistic case we employ a variational principle to obtain
\begin{lemma}\label{lemma:chupper}
For sufficiently large $Z$
$$
 E^C_\kappa(Z)  \leq 	\tr[ C_c(Z/|x|) d^C_<] +\tr[S(Z/|x|) d_>] + D(\rho_>,\rho_>) 
     + 2 D(\rho_<^C,\rho_>) + D(\rho^C_<,\rho^C_<).
$$
\end{lemma}

\begin{proof}
As noted above, $d^C$ satisfies $0\leq d^C\leq 1$ and $\tr d^C \leq Z$
for sufficiently large $Z$
\cite[Corollary~4.1]{SiedentopWeikard1987O}. Using that the
Hartree-Fock functional bounds the ground state energy from above --
even if non-idempotent density matrices are inserted, a fact that was
proven by Lieb \cite{Lieb1981} (see also Bach \cite{Bach1992}) -- and
estimating the indirect part of the Coulomb energy by zero we obtain
\begin{equation}
 E^C_\kappa(Z) \leq \tr[ C_c(Z|x|^{-1}) d^C ] + D(\rho^C,\rho^C).
\end{equation}
Both terms on the right-hand side are split according to $ d^C=d_<^C+d_>$.  To
obtain the desired upper bound we use the inequality $\frac 12 p^2 \geq
\sqrt{c^2p^2+c^4}-c^2$ for large angular momenta.
\end{proof}

The following lemma shows the irrelevance of the interaction energy of
the low lying states with all other electrons (including
themselves). The proof follows the strategy pursued in
\cite{SiedentopWeikard1987O}, namely to estimate it by the lowest
Coulomb energy of a particle in the field of an external point charge
$Z$, and then simply multiplying by the particle number. There is,
however, one important change in the channel $l=0$. Because of the
singular nature of the lowest eigenfunctions, their expectations in
potentials with Coulomb singularities does not exist. To circumvent
this problem we use the Hardy-Littlewood-Sobolev inequality followed
by a recent Sobolev-type inequality \cite{Franketal2006}.

\begin{lemma}\label{coulombsmall}
One has  $ D(\rho_<^C,\rho^C) \leq \const Z^{11/6}\log Z$.
\end{lemma}

\begin{proof}
We treat the terms $D(\rho_<^C,\rho_<^C)$ and $D(\rho_<^C,\rho_>)$
separately. For the latter one we recall that
\begin{equation}
  \label{eq:number}
  \int \rho^C_l(x) \, \rd x = (2l+1)(K-l), \qquad 0\leq l <L,
\end{equation}
where $K= O(Z^{1/3})$ and that by Proposition 3.4 in \cite{SiedentopWeikard1987O}
\begin{equation}\label{eq:potenergy}
\sum_{l=L}^\infty \int\frac{\rho_l(x)}{|x|} \rd x \leq \int \frac{\rho^S(x)}{|x|} \rd x \leq \const Z^{4/3}.
\end{equation}
The densities $ \rho_l^\# $ are spherically symmetric because of the
addition formula for the spherical harmonics.  Hence, using Newton's
theorem \cite{Newton1972}, we have
\begin{equation}
  \begin{split}
    D(\rho_<^C,\rho_>) &\leq \frac{1}{2} \int\rho_<^C(x) \rd x \int
    \frac{\rho_>(y)}{|y|}\rd y \\ &\leq \const
    \sum_{l=0}^{L-1}(2l+1)(K-l)Z^{4/3} \leq O(L^2 K Z^{4/3})
    =O(Z^{11/6}).
  \end{split}
\end{equation}
We set $\rho_{<>}^C := \rho^C_<-\rho^C_0$ and estimate 
\begin{equation}
D(\rho_<^C,\rho_<^C)\leq 2 D(\rho_0^C,\rho_{0}^C) + 2 D(\rho_{<>}^C,\rho_{<>}^C).
\end{equation}
This allows to treat the contributions from $l=0$ and $1\leq l< L$
separately. Using a scaled version of Lemma~\ref{ambi} with $R_l :=
\left((l+\frac12)^2-4\kappa^2\right)/4\kappa$ we obtain for $1\leq l
<L$
\begin{align*}
  \tr (|x|^{-1} d_l^C) 
  & \leq \frac1{2Z} \tr[C_c(0) \,  d_l^C] + \tr(\chi_{\{|x|>R_l/c\}}|x|^{-1}d_l^C)\\
  & \leq \frac12 \tr(|x|^{-1} d_l^C) +  \frac c{R_l} \tr d_l^C, 
\end{align*}
where the last inequality used the fact that eigenfunctions of $ d_l^C
$ are also eigenfunctions of $ C_c(Z|x|^{-1}) $ with negative
eigenvalue. Hence, summing over $l$ and noting that $R_l^{-1}\leq
\const\, l^{-2}$,
$$ \int\frac{\rho_{<>}(y)}{|y|}\rd y = \sum_{l=1}^L \tr (|x|^{-1}
	d_l^C) \leq \const Z \sum_{l=1}^{L-1} l^{-2}\int\rho_{l}^C(x)
	\,\rd x.
$$
Thus by
\eqref{eq:number} and again by Newton's theorem
\begin{align*}  
  D(\rho_{<>}^C,\rho_{<>}^C) & \leq \frac{1}{2} \int\rho_{<>}^C(x) \,\rd x \int
\frac{\rho_{<>}(y)}{|y|} \,\rd y \\  & \leq \const K L^2 \ K
Z\log L \leq \const Z^{11/6} \log Z.
\end{align*}

Finally, we treat the term corresponding to $l=0$. By the
Hardy-Littlewood-Sobolev inequality (c.f. \cite{LiebLoss2001}) and by
H\"older's inequality
\begin{align*}
D(\rho_0^C,\rho_0^C) & \leq \const\, \|\rho_0^C\|^2_{6/5} = \const\,
\left(\int \Big(\sum_{n=1}^K |\psi^C_{n,0,0}(x)|^2 \Big)^{6/5} \,\rd x
\right)^{5/3} \\ & \leq \const\, K^{1/3} \left( \sum_{n=1}^K \int
|\psi^C_{n,0,0}(x)|^{12/5} \,\rd x \right)^{5/3} .
\end{align*}
Now we use the Sobolev-type
inequality \cite[Eq.~(2.8)]{Franketal2006})
\begin{equation}
  \label{eq:sobolev}
  \|u\|_{12/5}^2 \leq \const  \left\langle u ,(|p|-\tfrac2{\pi}
  |x|^{-1}) u \right\rangle^{1/2} \, \|u\|
\end{equation} 
where the first factor on the right-hand side is to be understood in
form sense.  Using that $|p| - \tfrac2{\pi} |x|^{-1} \leq c^{-1}
C_c(Z|x|^{-1}) + c $ and that $\psi^C_{n,0,0} $ is a normalized
eigenfunction of $C_c(Z|x|^{-1})$ we deduce
\begin{equation}
\|\psi^C_{n,0,0}\|_{12/5} \leq \const c^{1/4} . 
\end{equation}
Combining the previous relations we arrive at
\begin{equation}
D(\rho_0^C,\rho_0^C) \leq \const\, K^{1/3}  (K c^{3/5} )^{5/3} \leq \const \, Z^{5/3}.
\end{equation}
This completes the proof of the lemma.
\end{proof}

\begin{proof}[Proof of Theorem~\ref{t1} -- second part]
It follows from Lemma \ref{coulombsmall} that 
$$ 2D(\rho_<^C,\rho_>) + D(\rho^C_<,\rho^C_<)=O(Z^{11/6}\log Z). $$
Hence Lemma~\ref{lemma:chupper} together with \eqref{eq:schlower} implies 
\begin{align}
    & \liminf_{Z\to\infty}Z^{-2} \left[ E^S(Z) - E^C_\kappa(Z) \right] \notag
	\\ & \qquad \geq \liminf_{Z\to\infty}Z^{-2}\bigg\{\tr\left[S(Z|x|^{-1})
	\, d^S_<\right] -\tr\left[ C_c(Z|x|^{-1}) \, d^C_< \right] \bigg\}
	\notag \\ & \qquad = \liminf_{Z\to\infty} \sum_{l=0}^{L-1} (2l+1)
	\sum_{n=1}^{K-l} Z^{-2} \left[ \left\langle \psi_{n,l,m}^S ,
	S(Z|x|^{-1}) \, \psi_{n,l,m}^S \right\rangle \notag \right. \\ &
	\mkern210mu \left. - \left\langle \psi_{n,l,m}^C , C_c(Z|x|^{-1})
	\, \psi_{n,l,m}^C \right\rangle \right] . \notag
\end{align}
The claim now follows from the scaling $x\mapsto x/c$ and Fatou's lemma.
\end{proof}

In order to get an explicit remainder estimate one could bound the sum
\begin{equation}
\sum_{l=0}^{L-1} (2l+1)
	\sum_{n=K-l+1}^\infty \left[ \left\langle \psi_{n,l,m}^S ,
	S(Z|x|^{-1}) \, \psi_{n,l,m}^S \right\rangle - \left\langle \psi_{n,l,m}^C , C_c(Z|x|^{-1})
	\, \psi_{n,l,m}^C \right\rangle \right]
\end{equation}
from above. This is certainly not difficult but for brevity we refrain
from doing so. The sum corresponding to $l\geq L$ can be bounded using
Theorem \ref{t:3}.



\end{document}